\documentclass{sn-jnl}

\jyear{2022}%

\usepackage{amsmath}
\usepackage{graphicx}%
\usepackage{amsfonts}%
\usepackage{amssymb}
\usepackage{natbib}
\usepackage{float}
\usepackage{multirow}

\theoremstyle{thmstyleone}%
\theoremstyle{thmstyletwo}%

\theoremstyle{thmstylethree}%

\newcommand{\bbeta}{ \mbox{\boldmath $ \beta $} }

\newcommand{\balpha}{ \mbox{\boldmath $ \alpha $} }

\newcommand{\bmu}{ \mbox{\boldmath $\mu$} }
\newcommand{\bOmega}{ \mbox{\boldmath $\Omega$} }

\newcommand{\bSigma}{ \mbox{\boldmath $\Sigma$} }

\newcommand{\bPsi}{ \mbox{\boldmath $\Psi$} }

\newcommand{\bgamma}{ \mbox{\boldmath $\gamma$} }

\newcommand{\boeta}{ \mbox{\boldmath $\eta$} }

\newcommand{\bzero}{\textbf{0}}
\newcommand{\bone}{\textbf{1}}

\newcommand{\bB}{\textbf{B}}

\newcommand{\bD}{\textbf{D}}

\newcommand{\bE}{\textbf{E}}

\newcommand{\bK}{\textbf{K}}

\newcommand{\bT}{\textbf{T}}

\newcommand{\bU}{\textbf{U}}

\newcommand{\bx}{\textbf{x}}
\newcommand{\bX}{\textbf{X}}
\newcommand{\by}{\textbf{y}}

\newcommand{\bR}{\textbf{R}}
\begin{document}

\title[Joint Modeling of Plants Traits and Reflectance]{Joint Multivariate and Functional Modeling for Plant Traits and Reflectances}

%\title[Joint Modeling of Plants Traits and Reflectance]{Joint Modeling of Plant Traits and Reflectances reveals Differing Dependence across Environments}

%Joint Modeling of Plant Traits and Reflectances with Application to the %Cape Floristic Region}
% \author*[1]{\fnm{Philip A.} \sur{White} \email{pwhite@stat.byu.edu}}
% \author[2]{}
% \author[3]{\fnm{Henry} \sur{Frye} \email{ADD}}
% \author[2]{\fnm{Alan E.} \sur{Gelfand} \email{ADD}} 
% \author[3]{\fnm{John A.} \sur{Silander, Jr. } \email{ADD}}

\author*[1]{\fnm{Philip A.} \sur{White} \email{pwhite@stat.byu.edu}}
\author[2]{\fnm{Michael F.} \sur{Christensen} \email{michael.f.christensen@duke.edu}}
\author[3]{\fnm{Henry} \sur{Frye} \email{henry.frye@uconn.edu }}
\author[2]{\fnm{Alan E.} \sur{Gelfand} \email{alan@duke.edu }} 
\author[3]{\fnm{John A.} \sur{Silander, Jr. } \email{john.silander\_jr@uconn.edu }} 

\date{}

% \thankstext{t2}{First supporter of the project}
%\thankstext{t3}{Second supporter of the project}
%\runauthor{P. White et al.}
\affil[1]{\orgdiv{Department of Statistics}, \orgname{Brigham Young University}, \city{Provo}, \postcode{84602}, \state{Utah}, \country{USA}}
\affil[2]{\orgdiv{Department of Statistical Science}, \orgname{Duke University}, \city{Durham}, \postcode{27708}, \state{Utah}, \country{USA}}
\affil[3]{\orgdiv{Department of Ecology and Evolutionary Biology}, \orgname{University of Connecticut},  \city{Storrs}, \postcode{06269}, \state{Connecticut}, \country{USA}}

\abstract{
The investigation of leaf-level traits in response to varying environmental conditions has immense importance for understanding plant ecology. Remote sensing technology enables measurement of the reflectance of plants to make inferences about underlying traits along environmental gradients. While much focus has been placed on understanding how reflectance and traits are related at the leaf-level, the challenge of modelling the dependence of this relationship along environmental gradients has limited this line of inquiry. Here, we take up the problem of jointly modeling traits and reflectance given environment. Our objective is to assess not only response to environmental regressors but also dependence between trait levels and the reflectance spectrum in the context of this regression. This leads to joint modeling of a response vector of traits with reflectance arising as a functional response over the wavelength spectrum. To conduct this investigation, we employ a dataset from a global biodiversity hotspot, the Greater Cape Floristic Region in South Africa.
}


\maketitle

\section{Introduction}\label{sec:intro}

%1. Introduction - \textbf{Henry, John - Would you both take a shot at filling in this intro a bit?  We have listed all of the points we think we need to make but not necessarily in the best sequence, not with the ecological importance, and certainly with no refs.  I assume you can find some off-the-shelf- material to insert.  You can also steal from the AoAs refelctance paper and even back to the Schliep et al. traits paper.  }

Terrestrial ecosystems are reliant on the diversity and composition of plant species present in a community \citep{oconnor_general_2017}.  Each species is comprised of a unique set of traits; these phenotypic characteristics include leaf size and shape, photosynthetic function, water content of leaves, and nutrient levels in leaves. The study of plant traits can provide insight into organismal function, plant–environment interactions, species coexistence and community dynamics, ecosystem structure and function, and biogeography and diversification. 
Thus, information on the diversity and abundance of plant traits can help us understand and predict complex ecological processes and responses to global change \citep{reich_tropics_1997, diaz_vive_2001, cadotte_beyond_2011}. Ecologists are particularly interested in how plant traits vary along spatial-environmental gradients since this can yield insights into the underlying principles of how communities originate and the convergence of survival strategies that plants evolved in response to their environment \citep{reich_generality_1999, mcgill_rebuilding_2006}.
These relationships provide a way to infer how ecosystems may change under novel environments, an important need as the rate human-driven environmental change increases \citep{schleuning_trait_based_2020}.

%Increasing interest in explaining and predicting trait behavior over space.  Growing literature on leaf-level reflectance. Studying reflectance at the leaf level is important because...   Well-established connection between traits and reflectances, with strength of association varying across reflectance bandwidth.  Dependence within reflectance, dependence between traits, dependence between traits and reflectance.  Environmental predictors influence trait levels and reflectances observed across locations. 

Remote sensing provides an efficient and powerful way to measure plant traits and diversity at regional extents \citep{turner_sensing_2014, cavender-bares_integrating_2022}. Image spectroscopy, which measures reflectance at high wavelength resolution across the reflectance spectrum, has been a prominent tool in predicting plant traits from remotely sensed imagery \citep{asner_taxonomy_2011,singh_imaging_2015, shiklomanov_quantifying_2016, yang_seasonal_2016}. At the leaf level, various wavelengths can be useful in predicting a suite of chemical, structural, and physiological traits \citep[see][and references therein]{jacquemoud_leaf_2019} and the diversity of leaf-level reflectance within a community has been shown to be correlated with community plant diversity \citep{schweiger_plant_2018, frye_plant_2021}. Further, leaf-level spectra provide the basis for understanding mechanisms occurring at larger scales observed by remote sensing instruments flown aerially and in space. 
%Given the range of traits across the world’s biomes, there 

There are still many gaps in our understanding of the relationship between plant optical properties and traits \citep{schimel_observing_2015, jetz_monitoring_2016}.
One of these gaps is an explicit understanding of how traits and reflectance respond in tandem to environmental conditions. Because the relationship between reflectance and traits is an important crux of modern remote sensing efforts, their dependency over different environments is a major question to be explored. However, the high dimensionality of spectral data alongside the co-correlation of individual wavelengths and leaf traits poses a methodological hurdle.
%for many ecologists interested in exploring such questions with empirically-based approaches\citep{kothari_plant_2022}. 
We offer novel joint modeling of traits and reflectances given environmental/habitat features. We consider joint modeling of two data types: multivariate continuous traits and functional reflectance data obtained at high wavelength resolution. We focus on understanding the effects of environmental regressors on plant traits and the reflectance spectrum at leaf-level, as well as the relationships between traits and reflectance, captured through correlations, under this regression. Modeling such relationships requires a multivariate and functional response to regressors, as well as a model that relates these responses. Traits may, in fact, be ordinal or categorical, e.g., the degree or state of leaf pubescence or waxiness, but consideration of such traits is beyond our scope here. 

%We offer novel joint modeling of traits and reflectances given environmental/habitat features. Here, we focus on joint modeling of two data types, i.e., multivariate continuous traits and functional reflectance data. That is, in our analysis, we focus on understanding the effects of environment on plant traits and the reflectance spectrum at leaf level, as well as the relationships between traits and reflectance, captured through correlations. Modeling such relationships requires multivariate and functional regression coefficients, as well as a model that relates multivariate trait to functional reflectance data. Traits may, in fact, be ordinal or categorical but we not considered such traits here \citep. 

%Following the above, the critical issue we address here is the strength of association between traits and reflectance and how such association varies across the reflectance spectrum/bandwidth. 

Conceptually, we can build trait/reflectance models over different taxonomic scales, e.g., family, genus, species. Here, we work at family scale to obtain the largest sample size of trait/reflectance data. This is needed in order to best understand the very large number of correlations of interest, i.e., four traits by $500$ wavelength bands, across $O(10^2)$ sites, each with individual environmental features. Thus, replicates are simply all of the observations available for the family in our database.
%Following the above, the critical issue we address here is the strength of association between traits and reflectance and how such association varies across the reflectance spectrum/bandwidth.  Conceptually, we can build trait/reflectance models over different taxonomic scales, e.g., family, genus, species.  Here, we work at family scale in order to obtain the largest sample size of trait/reflectance data.  This is needed in order to learn well the very large number of correlations of interest, i.e., four traits by $500$ reflectances, across $O(10^2)$ sites, each with individual environmental features. Thus, replicates are simply all of the observations available for the family in our database.   
%Joint modeling of families through, e.g., family level random effects, is not suitable because this implies a ``global centering'' of the families and borrowing strength across families.  Exploratory analysis below suggests that, in our context, trait behavior across families is sufficiently different so that such shrinkage is not meaningful, that global parameters over families, are not meaningful. Furthermore, different families have very different numbers of genera and species, further complicating joint interpretation.  Again, while analysis at a higher taxonomic scale is not appropriate with our data, if possible, our approach is applicable at higher scale and would enable potentially richer dependence stories.  

We remark that joint modeling of families, e.g., family level random effects, is not suitable because this implies a “global centering” of the families and borrowing strength across families. Exploratory analysis below suggests that, in our context, trait behavior across families is sufficiently different so that such shrinkage is not appropriate and that global parameters over families are not meaningful. Furthermore, different families have very different numbers of genera and species, further complicating joint interpretation. While analysis at a higher taxonomic scale is not appropriate with our data, our approach is applicable at higher scale and would enable potentially richer dependence stories.

We acknowledge that spatial dependence across sites is anticipated in joint modeling \citep{white2021spatial}, where, with different intentions, \cite{white2021spatial} focused on a marginal functional data model in a spatial setting. However, in attempting to obtain the needed large sample sizes for each of the families, we span regions that are disjoint and too spatially distant to employ sensible spatial modeling specifications. So, the observations are assumed to be conditionally independent. A fully spatial version, applied to an appropriate region, is a goal of future work. 

%We acknowledge that spatial dependence across sites is anticipated in joint modeling \citep{white2021spatial}.  However, in attempting to obtain large sample sizes for each of the families, we span regions that are disjoint and too spatially distant to employ sensible spatial modeling specifications. So, the observations are assumed to be conditionally independent.  A fully spatial version, applied to an appropriate region is a goal of future work.

Our primary contribution is jointly modeling a trait vector, $\bT$, and a functional reflectance spectrum, $\bR$, given environment/habitat features, $\bE$. Specifically, our model uses a joint multivariate and function-on-scalar regression, after which we can extract the residual association between $\bT$ and $\bR$ given $\bE$.  %We note that the joint response model here is for vector of traits with reflectance, as a functional response. In this regard,
We prefer a joint specification in the form $[\bT, \bR\vert \bE]$ to conditional times marginal specification, $[\bT \vert \bR,\bE][\bR \vert \bE]$ since the former directly reveals how we capture association in the residuals between traits and reflectances at the replicate level. Specifically, we directly model the correlation between $\bR$ and $\bT$ through a joint model for the coefficients of functional bases and trait residuals. Our model also provides functional heterogeneity and heteroscedasticity. 
%With different intentions, \cite{white2021spatial} focused on the marginal functional data model, $[\bR \vert \bE]$, in a spatial setting.

Turning to model assessment, under our specifications, we are assuming dependence among the traits, dependence across the reflectance spectrum, and dependence between traits and reflectances.  So, model comparison should be based upon conditional prediction.  We demonstrate improved out-of-sample prediction under the dependence model vs. an independence model given partial information at a site, i.e., when predicting traits or reflectances which we didn't collect at the site.

Multivariate modeling in ecology is well established \citep[see, e.g., ][]{schliep2013multilevel,clark2017generalized}. In particular, modeling the joint patterns of plant traits improves prediction \citep[see][]{schliep2018assessing}. Functional data analysis (FDA) is widely used to represent curves with continuous domains \citep[see][for pioneering work in the field]{ramsay2005,ramsay2007applied}. In general, FDA relies on representing the function through a low-rank represention (e.g., splines, wavelets, or kernels). Our challenge is relating multivariate data response specifications for traits to functional data response specifications for reflectance to allow relational inference between the responses.

We use our modelling framework to elaborate upon the joint $\bT$, $\bR$, and $\bE$ relationships for four families within the Greater Cape Floristic Region (GCFR) of South Africa. The GCFR is of special importance to global biodiversity as it contains two adjacent global  biodiversity hotspots, the Fynbos and Succulent Karoo biomes. Such biodiversity hotspots are important to global biodiversity conservation because they are regions that not only contain large numbers of species, but also have many species that are not found anywhere else on Earth \citep{myers_biodiversity_2000, latimer_neutral_2005, born_greater_2006}. The four families we focus our analysis on- the Aizoaceae, Asteraceae, Proteaceae, and Restionaceae- are iconic families found throughout the GCFR and are comprised of large species radiations \citep{manning_plants_2012, snijman_extra_2013}. We focus on four leaf traits- leaf water content, leaf mass per area, percent nitrogen, and succulence (water content/leaf area)- as these are both commonly used traits in the trait ecology and remote sensing literature and represent major evolutionary strategies among plants \citep[see][and references therein]{wright_worldwide_2004, jacquemoud_extraction_2019}.

%\textbf{Henry, John - Here, words just describe the stories of interest, stories that we elaborate in Section XXXx below -  environmental story across families, limitations/capability of reflectance data, issues with averaging over genera or species within families, dependence emerges as strongest at the individual plant level, i.e., individuals are viewed as the replications with regard to dependence, structured vs. unstructured residuals, interpreting association between traits and reflectance spectrum, environment doesn't seem to change the T and R association.  Perhaps the cleanest way to report the results of the analysis is to have a separate section for each family?  Then for each family, you can go through the E story as well as the dependence stories.}

%\textbf{Henry, John - Some lit review.  Some discussion regarding traits of interest.  Some discussion regarding leaf-level reflectance (you can steal for the AoAS paper). Some review of what is known about connection between traits and certain portions of the reflectance spectrum.}\\

%\textbf{A brief paragraph on the dataset.  Just one brief paragraph here, three or four sentences.  The exploratory data analysis follows in the next section.}\\

%We continue the paper with a presentation of the dataset and exploratory data analysis that motivates our analysis in Section \ref{sec:data}. Based on data characteristics presented in our exploratory data analysis, we present a joint model for multivariate traits and functional reflectance data in Section \ref{sec:model}. 

We continue the paper with a presentation of the dataset and exploratory data analysis that motivates our analysis in Section \ref{sec:data}. Based on data characteristics demonstrated in our exploratory data analysis, we present a joint model for multivariate traits and functional reflectance data in \ref{sec:model}. We then offer interpretation of our results for each of the four families in Section \ref{sec:results} and conclude with a brief summary and potential future work.

\section{The Dataset}\label{sec:data}

\begin{figure}[h!]
    \centering
\includegraphics[width = \textwidth]{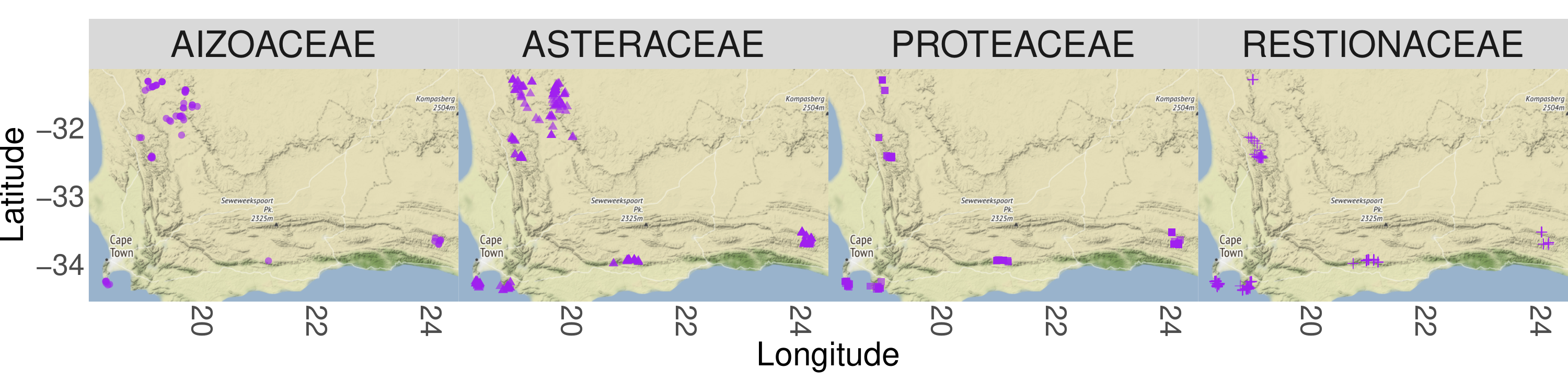}

%\vspace{-4mm}

    \caption{Locations of trait and reflectance data for each plant family.}
    \label{fig:locations}

\end{figure}

We model plant trait and reflectance data at the family level, fitting each of the families Aizoaceae, Asteraceae, Proteaceae, and Restionaceae individually. These four families are generally speciose in the Greater Cape Floristic Region (GCFR) in South Africa, though the Proteaceae and Restionaceae have less prevalence in the arid regions of the GCFR \citep{manning_plants_2012, snijman_extra_2013}. In Table \ref{tab:counts}, we provide the number of times each family is observed in the dataset as well as the number of genera and species within each family. We see that the families differ dramatically in how often they appear, as well as how many genera and species appear for that family. Although some plant families have similar spatial ranges, they are not generally co-located (See Figure \ref{fig:locations}).

\begin{table}[h!]
\centering
    \caption{Number of observations and genera for each family considered in this study.}
    \label{tab:counts}
\begin{tabular}{rrrr}
  \hline
Family & Observations (N) & Genera & Species\\ 
  \hline
     Aizoaceae & 64 &  17 & 45 \\ 
   Asteraceae & 310 &  58 & 197 \\ 
   Proteaceae & 110 & 10 & 58 \\ 
  Restionaceae & 152 & 16 & 89 \\ 
   \hline
\end{tabular}

\end{table}

%We analyze four continuous plant traits -- fresh water content (FWC), leaf mass area (LMA), percent Nitrogen (pN), and leaf succulence (LS) along with reflectance as a function over wavelength $w \in [450,950]$ nanometers, observed as a $500$ dimensional reflectance vector, by nanometer. In some cases, at a site, for a given species within a family, we have more than one trait observation and/or reflectance observation. Again, viewing the sites as replicates, such observations are averaged to obtain a single $\bT$ value and a single $\bR$ value for the site. Thus, we do not have more than one observation for any species at a site.  However, there may be duplication of replicates at a site because more than one species is sampled at many sites. 

We analyze four continuous leaf traits – leaf water content (LWC), leaf mass per area (LMA), percent Nitrogen (pN), and leaf succulence (LS) along with leaf reflectance as a function over wavelength $w \in [450,950]$ nanometers, observed as a $500$ dimensional reflectance vector, by nanometer. Leaf reflectance was measured from sun leaves collected from the top of the canopy using a USB-$4000$ Spectrometer (Ocean Optics, Largo, Florida, USA) with a leaf clip attachment. For further details on spectra and trait data collection see \cite{frye_plant_2021} and \cite{aiello-lammens_processes_2017}, respectively. In some cases, at a site, for a given species within a family, we have more than one trait observation and/or reflectance observation. Again, viewing the sites as replicates, such observations are averaged to obtain a single $\bT$ value and a single $\bR$ value for the site.  However, there may be duplication of replicates at a site because more than one species is sampled at many sites.

We analyze traits and and reflectance on the log-scale as a standard transformation that improves the assumptions of our Gaussian model (discussed in Section \ref{sec:model} below). 
%We model the data at family level because there is often very limited replication at species or genus level. 
We define the response as $\by_{j} = \left( \bT_{j}' , R_{j}(w) \right)'$, where $\bT_{j}$ is a vector of four plant traits (on the log scale) and $R_{j}(w)$ is a log-reflectance function. We model $R_{j}(w)$ as a random function where, again, $R_{j}(w)$ is observed at 500 wavelengths at one nanometer (nm) spacing between 450-949 nm. For both traits and reflectance, we use $j$ to index replication within family. %\textbf{From Alan: I think we should add a little more detail about how these data are acquired at a given site.  Perhaps a short new paragraph?}

\begin{figure}[ht!]
    \centering
    \includegraphics[width = \textwidth]{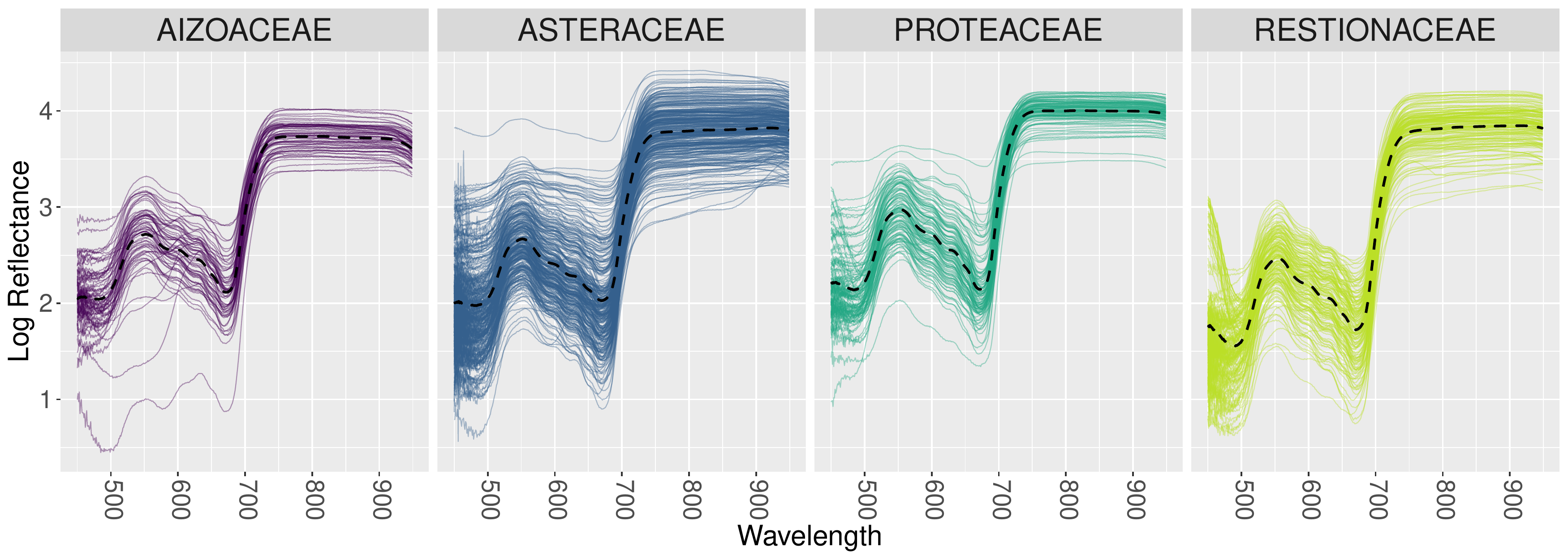}
\includegraphics[width = \textwidth]{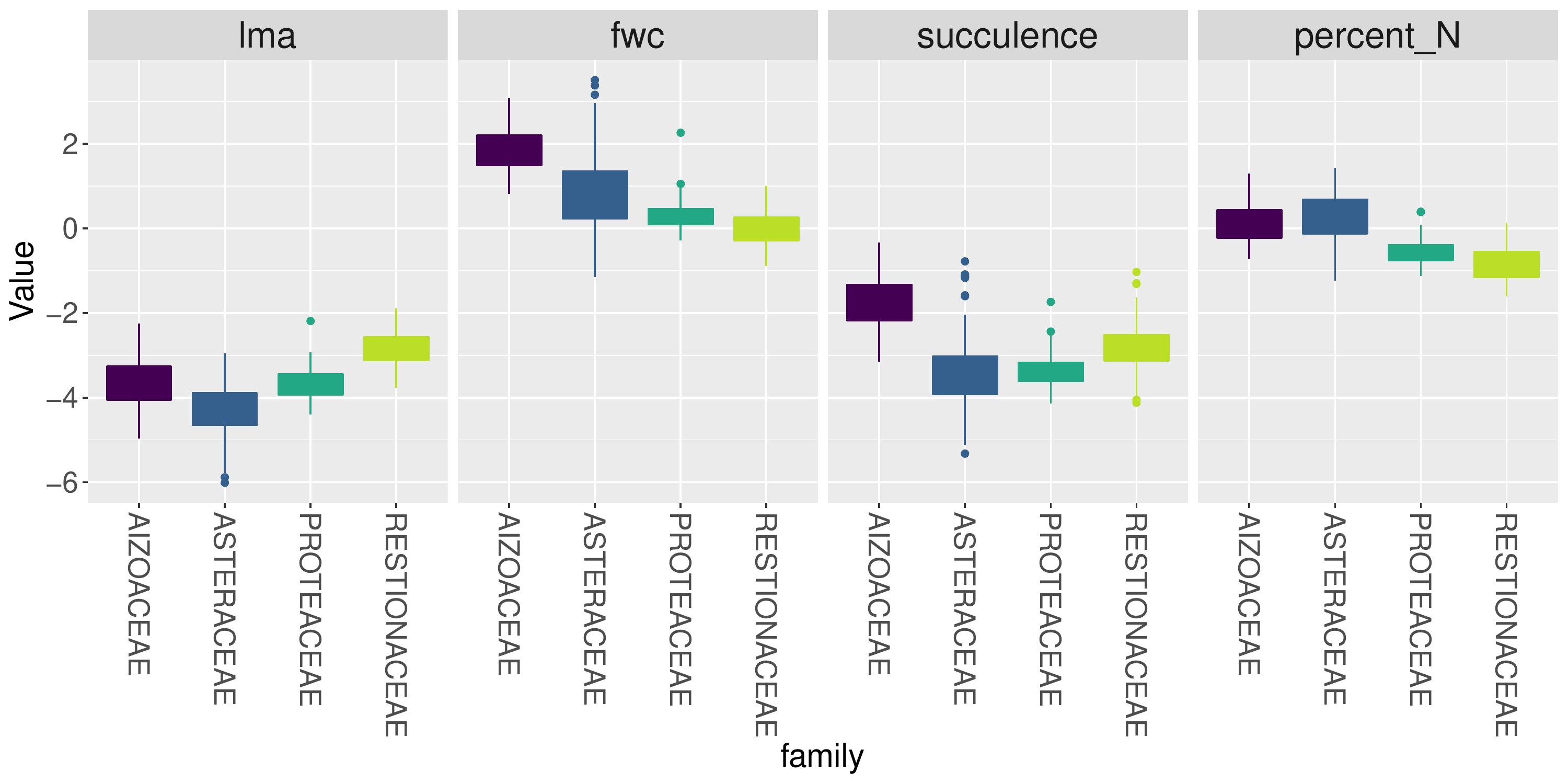}
    \caption{(Top) All reflectance curves for each family. Dashed line represents the family-specific mean. (Bottom) Boxplots for the four plant traits, paneled by each family.}
    \label{fig:family_EDA}
\end{figure}

To visualize the overall patterns present in the data, we plot all log reflectances by family in Figure \ref{fig:family_EDA}, including the family-specific mean. All families show some similarities with relatively low reflectance for blue (450 - 500 nm) and some red wavelengths (600 - 675 nm), with a local maximum around 550 nm (green). Reflectance increases in the red (around 700 nm) and remains uniformly high in the near-infrared (740-949 nm). In Figure \ref{fig:family_EDA}, we also include box plots of the observed log plant traits for each family. Although traits and reflectances are similar across families, each family shows different amounts of heterogeneity. 
%\vspace{-4mm}

%\vspace{-4mm}

To jointly explain plant traits and reflectance, we use four environmental covariates in $\bE_{j}$: (i) elevation (Elevation30m), (ii) annual precipitation (Gmap), (iii)rainfall concentration (RFL\_CONC), and (iv) minimum average temperature in January, the peak of the austral summer (tminave 01c). Elevation data was derived from 30 m resolution digital elevation maps \citep{nasa_jpl_nasadem_2020} while the other climate variables were taken from \cite{schulze_south_1997}. We also introduce the family-level plant abundance at the site of the $j$th replicate as an explanatory variable.  Though not a customary environmental predictor, we include it in $\bE_j$ to play the role of a proxy for site level environmental suitability for the family. That is, the other environmental regressors above operate at larger spatial scales and we seek to supply a more local regressor.  

As a measure of suitability, we define abundance for a family at a site as the aggregated percent cover of all species in that family at the site. Due to some misalignment between the sites in this analysis and sites with available percent cover, we estimate percent cover using ordinary kriging on the scale of $\log(x +1)$ to yield predictions on $[0,\infty)$ as well as to deal with many zeros. Specifically, we use an exponential covariance function with parameters estimated from empirical semivariograms.

\begin{figure}[h!]
    \centering
    \includegraphics[width = \textwidth]{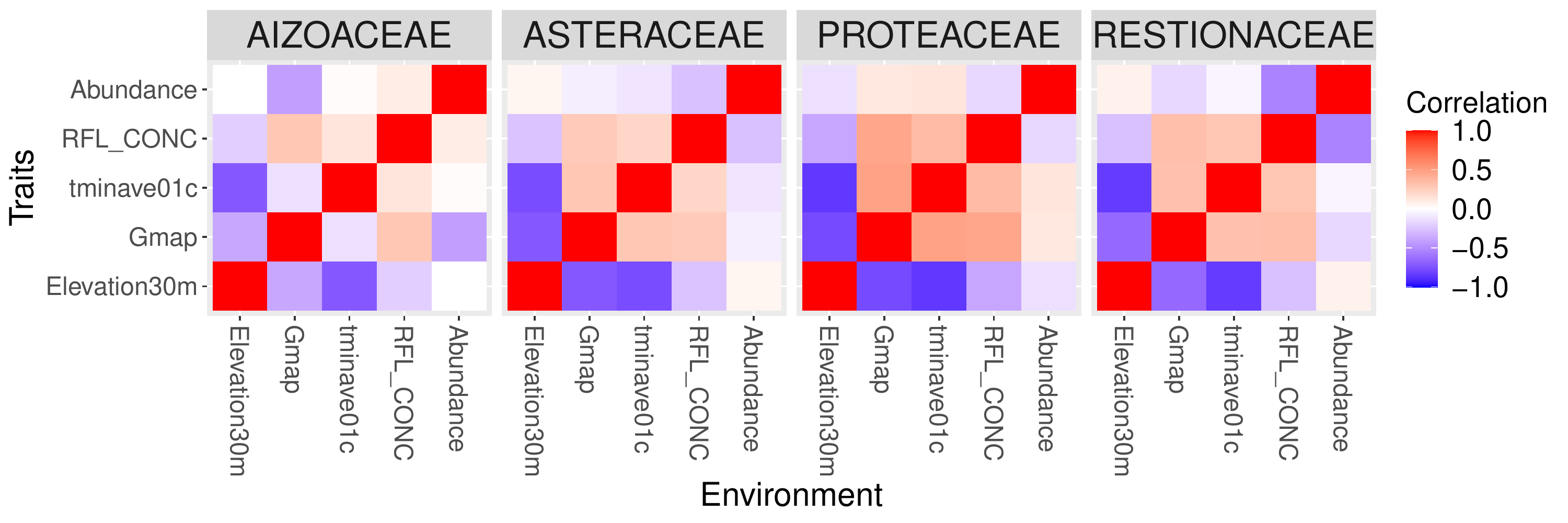}
\includegraphics[width = \textwidth]{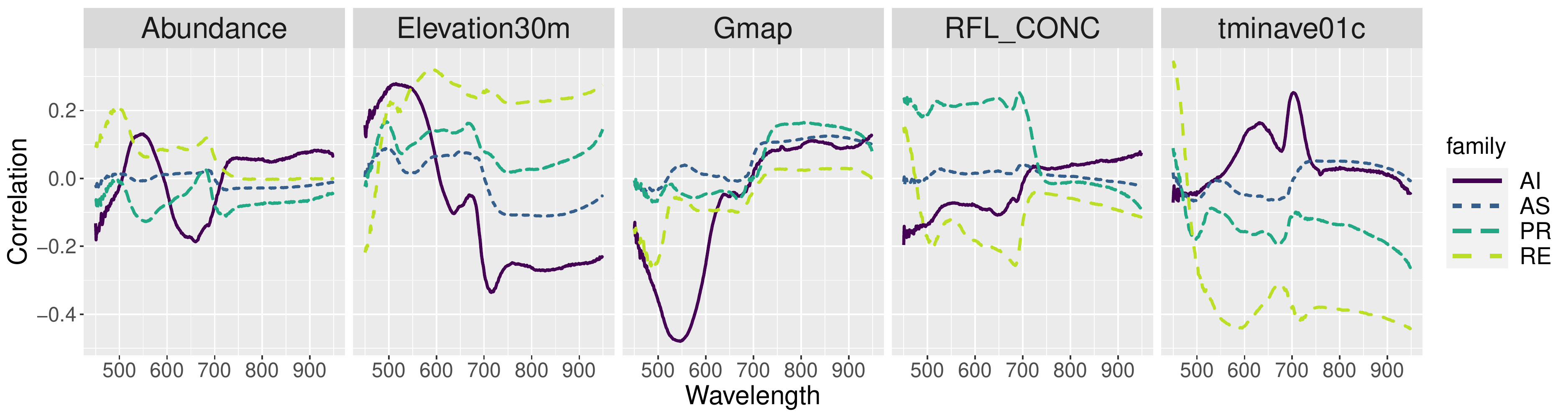}
\includegraphics[width = \textwidth]{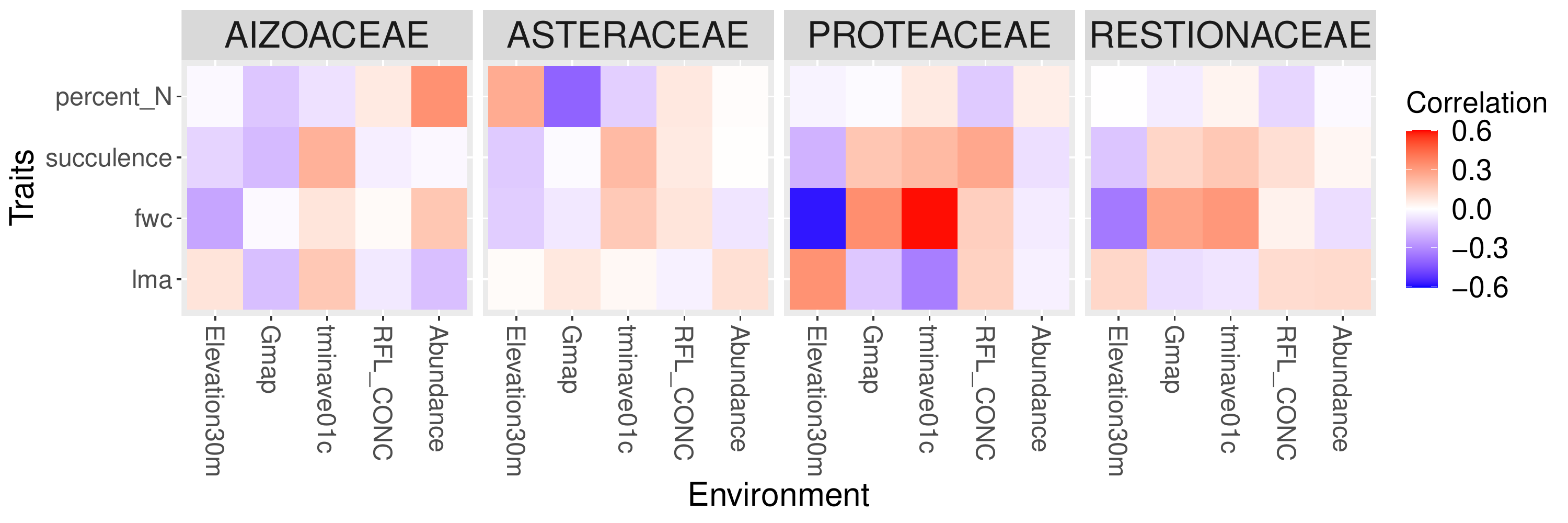}
    \caption{Empirical correlations between environmental predictors (Top) Themselves, (Middle) log traits, and (Bottom) log reflectance. The results are presented for each plant family.  The labels ``AI'', ``AS'', ``P'', and ``R'' represent Aizoaceae, Asteraceae, Proteaceae, and Restionaceae, respectively.}
    \label{fig:envr_EDA}
\end{figure}

To motivate our analysis, we calculate the empirical correlation between the environmental variables and each log trait and log reflectance for each family. We plot these correlations in Figure \ref{fig:envr_EDA}. 
Apart from a few exceptions, most correlations between $\bE_j$ and $\bT_j$ are weak, not surprising given only modest correlations between traits and environment within the region \citep{mitchell_functional_2015,aiello-lammens_processes_2017} and at global and other local scales \citep{wright_worldwide_2004, wright_does_2012}. Importantly, there is very little common correlation pattern shared across families, expected given the large differences in growth form and likely ecological strategies that each lineage has evolved. As with traits, examining the empirical correlations between the environmental variables and the log reflectance curves reveals little common pattern between the families. However, importantly, there are evident differences in the correlations between reflectance and environment over the wavelength spectrum, suggesting the need for wavelength-varying coefficients in functional modeling of reflectance as a response.

%As with traits, examining the empirical correlations between the environmental variables and the log reflectance curves reveals little common pattern between the families. Importantly, there are evident differences in the correlations between reflectance and environment over the wavelength spectrum, suggesting the need for wavelength-varying coefficients in functional modeling of reflectance as a response.

%Because our primary target is identifying relationships between traits and reflectance after accounting for environment, 

\begin{figure}[h!]
    \centering
\includegraphics[width = \textwidth]{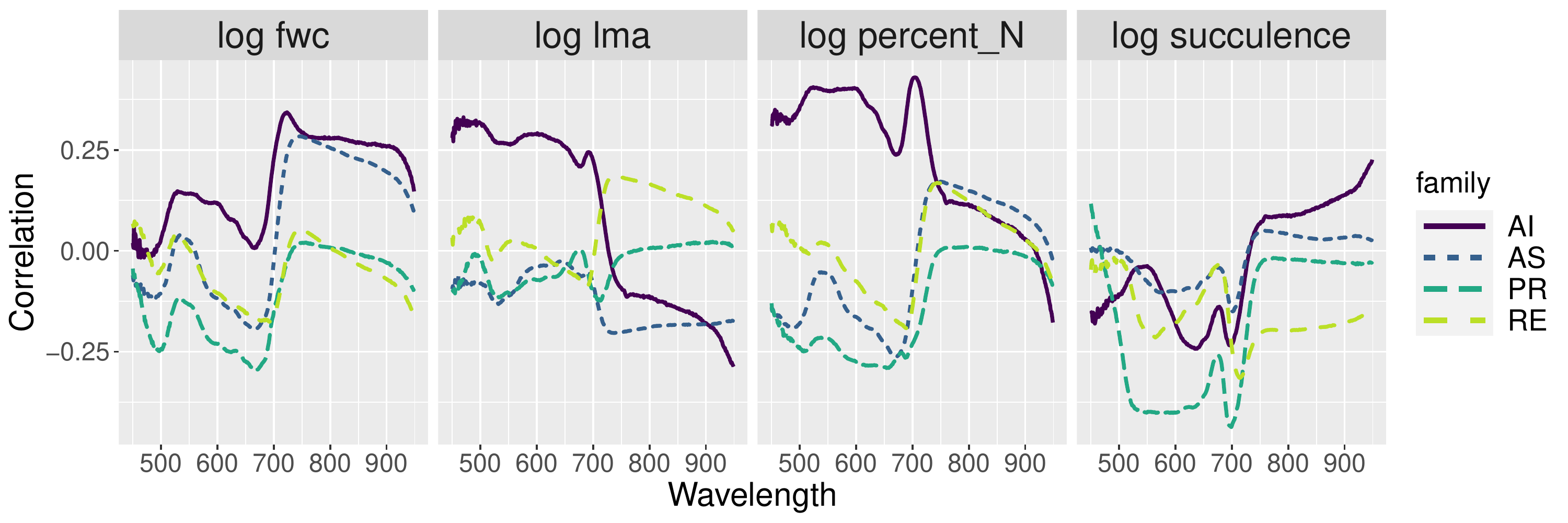}
    \caption{Empirical correlations between traits and log reflectances for each family. The labels ``AI'', ``AS'', ``P'', and ``R'' represent Aizoaceae, Asteraceae, Proteaceae, and Restionaceae, respectively.}
    \label{fig:trait_reflect_EDA}
\end{figure}

We present the empirical correlations between log traits and log reflectance in Figure \ref{fig:trait_reflect_EDA}. 
There are few similarities in the trait-reflectance correlations among the families. In general, the families with more data have weaker correlations between traits and reflectances. Families with more genera and species likely represent speciation “hot-beds”, e.g., see \cite{verboom_origin_2009}, \cite{pirie_biodiversity_2016}, and \cite{mitchell_anchored_2017}, where we would expect that the number of species would result in greater variability of traits and reflectance within lineages. It has been shown that higher within-group trait variation dilutes trait associations \citep{laughlin_intraspecific_2017, anderegg_withinspecies_2018} and thus we may anticipate a similar effect for trait and reflectance relationships. Again, our goal is to assess the strength of these trait/reflectance relationships while accounting for environment.

Our exploratory analysis shows that the relationships between traits, reflectance, and environmental predictors (See Figures \ref{fig:envr_EDA} and \ref{fig:trait_reflect_EDA}) differ greatly across families, both in shape and magnitude. 
%Moreover, many of the families are not co-located. 
Further, in preliminary modeling efforts, we found no benefit to modeling the families jointly. 
%Altogether, these issues raise the question of the appropriateness of modeling these families jointly. 
Therefore, as noted above, we model each family separately.

\section{Model and Methods}\label{sec:model}

\subsection{The joint specification}

%As discussed, we model each plant family separately because our exploratory analyses do not suggest common data patterns across family. 

Since we model the families individually, we need only subscript the replicates within a family.  So, consider vector $\bT_{j}$, an $ s \times 1 $ vector of  trait responses for replicate $j$.  We assume, after the log-transformation, that this vector can be modeled to follow a multivariate normal distribution. With a binary or categorical response we would  view the corresponding entry in $\bT_{j}$ as \emph{latent}, driving the observed response.  We model $\bT_{j}$ as
\begin{equation}
\bT_{j} = \balpha^{(T)} + \bB^{(T)}\bE_{j} + \bU_{j}^{(T)}.
\end{equation}
Here, $\balpha^{(T)}$ are trait-specific intercepts, $\bE_{j}$ is a vector of environmental predictors, say $p \times 1$, for replicate $j$ and $\bB^{(T)}$ is a $s \times p$ matrix of trait-specific regression coefficients.  The $\bU_{j}^{(T)}$ are pure errors, i.i.d. $\sim MVN(\bzero, \Omega^{(T)})$.

We model the reflectance as a functional response variable, observed at $500$ wavelengths, with wavelengths denoted by $w$'s.  Specifically,
\begin{equation}
R_{j}(w) = \alpha^{(R)}(w) + \bE_{j}^{'}\bbeta^{(R)}(w) + \bK_U^{'}(w)\bU_{j}^{(R)} + \psi_{j}(w).
\end{equation}
Here, $\alpha^{(R)}(w) = \bK_{\alpha}^{'}(w) {{\balpha^*}^{(R)}}$ is a wave-length varying intercept, where dimension reduction is given through $l$ basis functions in $\bK_{\alpha}^{'}(w)$. $\bE_{j}$ is, as above with the $p \times 1$ wavelength specific coefficient vector $\bbeta^{(R)}(w) \equiv \bB^{(R)}\bK_{\beta}(w)$.  That is, we imagine $\bK_{\beta}(w)$ as an $m \times 1$ vector of basis functions or convolution functions with $\bB^{(R)}$ an $p \times m$ matrix of coefficients providing dimension reduction.  
%The wavelength coefficient vectors are written through dimension reduction.  
Aggregating, $\bbeta_{p \times 500}^{(R)} = \bB^{(R)}\bK_{\beta}'$ where $\bK_{\beta}$ is $500 \times m$.% \textbf{Why don't you mention $\bK_{\alpha}^{'}(w)$?}

Further, again using dimension reduction, the term $\bK_U^{'}(w)\bU_{j}^{(R)}$ introduces the $q \times 1$ replicate level vector $\bU_{j}^{(R)}$ of reflectance random effects to supplement the fixed effects contribution.  These random effects vectors adopt $q << 500$ to provide a dimension reduction for the reflectances.  More will be said about the choice of $q$ below. The $\bK_{U}(w)$ can be collected into a $500 \times q$ matrix, $\bK_{U}$. Finally, the $\psi_{j}(w)$ provide independent wavelength specific pure error terms with variances $\sigma^{2}(w)$. We model $\log(\sigma^2(w))= \bK_\sigma \bgamma_\sigma$ as a linear spline as in \cite{white2021spatial}.
%\textbf{I still question introducing $500$ of these in a diagonal matrix but you seem to be OK with it!}
The entire reflectance response vector for replicate $j$ becomes $\bR_{j} = \balpha^{(R)} + \bK_{\beta}^{T}\bB^{(R)'}\bE_{j} + \bK_U\bU_{j}^{(T)} +\bPsi_{j}$.  %\textbf{I think we can keep the $\alpha^{(R)}$ vector here for now. If you wish to absorb it into the $\bB^{(R)}$ matrix later, fine with me.}

We introduce dependence between the traits and reflectances, the $T$'s and the $R$'s, through the $U$'s.  That is, the dependence is at the replicate level.  Specifically, we assume $\left(
                                                                                                  \begin{array}{c}
                                                                                                    \bU^{(T)}_{j} \\
                                                                                                    \bU^{(R)}_{j} \\
                                                                                                  \end{array}
                                                                                                \right)$ is distributed as a mean $\bzero$ multivariate normal
with covariance matrix $\bOmega = \left(
                          \begin{array}{cc}
                            \bOmega^{(T)} & \bOmega^{(TR)} \\
                            \bOmega^{(TR)'} & \bOmega^{(R)} \\
                          \end{array}
                        \right)$.  As a result the induced covariance matrix for $\left(
                                                                                    \begin{array}{c}
                                                                                      \bT_{j} \\
                                                                                      \bR_{j} \\
                                                                                    \end{array}
                                                                                  \right)$ becomes $\bSigma = \left(
                                                                                                                \begin{array}{cc}
                                                                                                                  \bOmega^{(T)} & \bOmega^{(TR)}\bK_U' \\
                                                                                                                  \bK_U\bOmega^{(TR)'} & \bK_U\bOmega^{(R)}\bK_U' + \bD_{\psi} \\
                                                                                                                \end{array}
                                                                                                              \right)$ where $\bD_{\psi}$ is the diagonal matrix of pure error reflectance variances.
From this matrix we can extract all covariances and correlations.  
%(\textbf{I guess you could insert some specific forms, i.e., trait-trait, reflectance-reflectance, trait-reflectance but perhaps not necessary.}  
If $\bOmega^{(TR)}$ is a matrix of zeros, then we have an independence model, which we denote as $[\bT\vert \bE][\bR\vert\bE]$. 

We provide details of the dimension reduction used in Section \ref{sec:dim_red}, justify the use of the joint model through cross validation in Section \ref{sec:mod_comp} and Appendix \ref{app:mod_comp}, and then present the results using the joint model in Section \ref{sec:results}.

\subsection{Dimension reduction details}\label{sec:dim_red}

%Because dimension reduction plays a key role in our functional specification of $R_j(w)$, we provide details here. 

Following \cite{white2021spatial}, we specify the low-rank functional terms wavelength-varying intercept $\alpha^{(R)}(w)$, random effects $ \bK_U^{'}(w) \bU_{j}^{(R)}$, and $\bbeta^{(R)}(w)$ through process convolutions enabling simple connection to Gaussian processes (GPs).  That is, the kernels of the process convolution connect the low-rank process to the GP covariance \citep{higdon2002space}. For every basis, we include an intercept so that the wavelength-varying intercepts, regression coefficients, and variances have an overall centering. 

We use a rich specification for the wavelength-varying intercept $\alpha^{(R)}(w)$, employing Gaussian kernels with wavelength knots spaced every 10 nm from 450-950 nm ($N_\alpha =52$, in total, including the intercept) to obtain $\bK_\alpha(w)$. To specify the wavelength-varying random effects through $\bK_U(w)$, we use Gaussian kernels with wavelength knots spaced every 25 nm from 450-950 nm ($N_U =22$, in total, including an intercept). To specify wavelength-varying coefficient functions, we use Gaussian kernels with wavelength knots spaced every 100 nm from 450-950 nm ($N_\beta =7$, including the intercept) for $\bK_\beta(w)$.  We allow the scale parameters for the dimension-reduced coefficient to be unknown.  However, we fix the bandwidth of the Gaussian kernels to be 1.5 times the kernel spacing to alleviate well-known lack of identifiability with scale and range parameters of Gaussian process models \citep[see, e.g.,][]{zhang2004inconsistent}.  Lastly, we use a linear splines with interior knots every 50 nm from 475-925 nm to specify $\bK_\sigma$ ($N_\sigma =12$, in total). The selection of knot spacing chosen here is motivated by a sensitivity analysis in \cite{white2021spatial}, where the simplest model is chosen that does not significantly decrease model performance. 

\subsection{Prior Distributions, Model Fitting, and Prediction}\label{sec:priors}

We adopt weakly informative prior distributions for all trait regression coefficients and the intercepts of the wavelength-varying parameters for reflectance. However, for wavelength-varying intercept and regression coefficient functions, we have unknown scale parameters that shrink the low-dimensional functional bases toward zero. Overall, these priors assume that the wavelength-varying parameters receive an overall centering given by associated intercepts. We use proper prior distribution with large variance for all variance and covariance parameters. Specifically, we use the following prior distributions:

\begin{equation}\label{eq:priors}
\begin{aligned}
{{\balpha^*_1}^{(R)}} &\sim N\left(0 ,10^3 \right),\\
{{\balpha^*_j}^{(R)}} &\overset{iid}{\sim} N\left(0 ,\sigma^2_\alpha \right); j = 2,...,N_\alpha, \\
{\bB}^{(R)}_{k1} &\overset{iid}{\sim} N\left(0,10^3 \right); k = 1,...,p;  \\
{\bB}^{(R)}_{kj} &\overset{iid}{\sim} N\left(0,\sigma^2_\beta \right); k = 1,...,p; j = 2,...,N_\beta, \\
\balpha^{(T)}_k &\overset{iid}{\sim} N\left(0,10^3 \right); k = 1,...,p, \\
{\bB}^{(T)}_{jk} &\overset{iid}{\sim} N\left( 0,10^3 \right); j = 1,...,s ;k = 1,...,p, \\
\end{aligned}
\qquad
\begin{aligned}
{\bgamma_\sigma}_{1}  &\sim N\left(0,10^4 \right)\\
{\bgamma_\sigma}_{j} &\overset{iid}{\sim} N\left(0,9 \right); j = 2,...,N_\sigma, \\
\bOmega^{-1} & \sim \text{Wishart}\left(s + N_U + 1, 10^{-3} \mathbb{I} \right),\\
\sigma^{-2}_\alpha &\sim \text{Gamma}(1,1),\\
{\sigma_\beta}^{-2}_k &\overset{iid}{\sim} \text{Gamma}(1,1)  ; k = 1,...,p,
\end{aligned}
\end{equation}

Letting $\boeta$ denote all model parameters, we estimate $\boeta$ using Markov chain Monte Carlo (MCMC). So, our model fitting yields $M$ samples from the posterior distribution of all model parameters $\boeta_1,..., \boeta_M$. For all parameters except $\bgamma_\sigma$, we are able to use a Gibbs sampler because posterior conditional distributions can be found in closed form. For $\bgamma_\sigma$, we use a Metropolis within Gibbs update with a Gaussian random walk proposal distribution. We run this algorithm for 200,000 iterations, discard a burn-in period of 100,000 iterations, and, to limit memory requirements, thin the remaining 100,000 samples to 5,000 samples. During the burn-in period of the model-fitting, we tune acceptance rate to be between 0.2 and 0.6 during the burn-in period of the model fitting. Specific details of the posterior conditional distributions are provided in Appendix \ref{app:gibbs}.

Beyond the primary goal of inferring relationships between environment, traits, and reflectance,  a further use for this model is prediction of traits or reflectance in the frequent scenario where, at a given site, measurements of either plant traits or reflectances were made but not both. In such settings, the residuals/random effects $U$ would be attached to only partially observed samples. Therefore, predictions for reflectance or traits are made by conditionally predicting U, $U^{(R)}\vert U^{(T)}$ or $U^{(T)}\vert U^{(R)}$, respectively. Under our model, these predictions rely on conditional normal theory. To illustrate this, we drop the $j$ subscript and consider prediction using a single posterior sample $\boeta_m$ of all model parameters. The conditional prediction of traits or reflectance would, respectively, be
\begin{equation*}
    \begin{aligned}
    \tilde{T}_m &= \balpha^{(T)}_m + \bB^{(T)}_m \bE + \bU_m^{(T)},\\
\tilde{R}(w) &= \alpha_m^{(R)}(w) + \bE^{'}\bbeta_m^{(R)}(w) + \bK_U^{'}(w)\bU_{m}^{(R)} + \psi_{m}(w),    
\end{aligned}
\end{equation*}
where 
 $\bU_m^{(T)} \sim N(\bmu_{T\vert R},\bSigma_{T\vert R}),$,  $\bmu_{T\vert R} = \bOmega^{(TR)} {\bOmega^{(R)}}^{-1} \bU^{(R)}_m,$
 $\bSigma_{T\vert R} = \bOmega^{(T)} -\bOmega^{(TR)} {\bOmega^{(R)}}^{-1} {\bOmega^{(TR)}}',$
  $\bU_m^{(R)} \sim N(\bmu_{R\vert T},\bSigma_{R\vert T}),$
   $\bmu_{R\vert T} = \bOmega^{(RR)} {\bOmega^{(T)}}^{-1} \bU^{(T)}_m,  $ and
$ \bSigma_{R\vert T} = \bOmega^{(R)} -{\bOmega^{(TR)}}' {\bOmega^{(T)}}^{-1} {\bOmega^{(TR)}}'.$
This process is repeated for all posterior samples $\boeta_1,..., \boeta_M$, yielding a set of $M$ predictions. 

\subsection{Model Comparison}\label{sec:mod_comp}

We justify the joint model specification $[T,R\vert E]$ by comparing conditional predictive performance using 10-fold cross-validation. Specifically, we compare the joint model to a model where traits and reflectance are independent $[T\vert E][R\vert E]$. For each fold of the cross validation, we hold out 10\% of all traits (jointly), as well as 10\% of reflectance spectra (the entire spectrum). The traits and reflectance spectrum are held out exactly one time in the 10-fold cross validation. We adopt this comparison approach to mirror the scenario above where either plant traits or reflectances are measured but not both.  Of course, other holdout schemes could be investigated.
%Thus, we are interested in our ability to predict missing plant traits or reflectances conditional on other. 
%With matching interest in conditionally predicting traits or reflectance, we weigh the ability of the models to predict reflectances and traits equally.  \textbf{Does it make sense to separate the cond'l prediction for T given R and R given T?  Do we do better with one than with the other?}

 We focus our comparison here on the Asteraceae family because it has the most data to help in estimating the proposed correlation structure. The model comparison results are summarized in Table \ref{tab:comp}, while the comparison results for the other families are in Appendix \ref{app:mod_comp}. We compare models by using predicted root mean squared error (RMSE), mean absolute error (MAE), and the mean energy score (ES),
 $$ES(F,\bx) = \frac{1}{2} \mathbb{E}_F \| \bX  - \bX' \| - \mathbb{E}_F \| \bX - \bx\|,$$
 where $\bX,\bX'$ follow the same distribution $F$ and $\bx$ is a vector of hold-out values \citep[see][]{gneiting2007strictly}. To estimate this empirically, from a set of posterior predictions $\bX_1,...,\bX_M$, forming $\hat{F}$ for a vector $\bx$, the energy score is calculated as
 $$ES(\hat{F},\bx) = \frac{1}{2M^2} \sum^M_{m = 1} \sum^M_{m' = 1} \| \bX_m - \bX_{m'} \|  - \frac{1}{M} \sum^M_{m = 1} \| \bX_m - \bx \|,$$
 and we average this for all hold-out vectors.
The ES compares multivariate predictions to multivariate quantities and is a proper scoring rule \citep{gneiting2007strictly}. We use the ES as our primary model selection criterion when predicting all traits or all reflectances. We present MAE and RMSE for each trait individually but, for simplicity, choose to average these quantities for reflectances over all wavelengths. On the other hand, because either traits or reflectances are predicted jointly conditioning on the other, a single mean ES is given for all traits and for all reflectances, respectively. 
%The model comparison results for the Asteraceae family are summarized in Table \ref{tab:comp}, while the comparison results for families with fewer observations in Appendix \ref{app:mod_comp}.

For the Asteraceae family, the joint model improves out-of-sample prediction performance for reflectances and all traits. Overall, the benefit of the joint model is much larger for reflectance than traits, and this benefit is substantial. Based on these findings, we use the joint model to present interpretation of the results. 
%We also find similar but weaker results for the other families; however, the joint model does not uniformly improve trait prediction for all families. \textbf{A bit of shooting ourselves in the foot!}
 
\begin{table}[h!]
\centering
\caption{Model comparison between joint and independent models for Asteraceae.}\label{tab:comp}
\begin{tabular}{|l|l|r|r|r|}
  \hline
Quantity & Model & MAE & RMSE & ES \\ 
  \hline  
log LMA & \multirow{4}{*}{$[T\vert E][R\vert E]$}  & 0.449 & 0.557 & \multirow{4}{*}{0.826} \\ 
  log FWC &  & 0.646 & 0.814 &  \\ 
  log LS &  & 0.533 & 0.706 &  \\ 
  log pN & & 0.395 & 0.495 & \\ 
  \hline
  log Reflectance & $[T\vert E][R\vert E]$ & 0.543 & 0.981 & 16.396 \\
  \hline \hline
  log LMA & \multirow{4}{*}{$[T,R\vert E]$} & 0.398 & 0.517 & \multirow{4}{*}{0.658} \\ 
  log FWC &  & 0.465 & 0.612 & \\ 
  log LS &  & 0.401 & 0.546 &  \\ 
  log pN &  & 0.343 & 0.457 &  \\ 
  \hline
  log Reflectance & $[T,R\vert E]$ & 0.166 & 0.244 & 3.703 \\ 
   \hline
\end{tabular}

\end{table}

% \begin{table}[h!]
% \centering
% \begin{tabular}{rrrrr}
%   \hline
%  &  Asteraceae & Restionaceae & Proteaceae & Anacardiaceae \\ 
%   \hline
% FWC & 0.0089 & -0.0031 & -0.0020 & -0.0128 \\ 
%   LMA & 0.0067 & 0.0270 & 0.0059 & 0.0098 \\ 
%   pN & 0.0162 & 0.0024 & 0.0087 & -0.0110 \\ 
%   LS & 0.0137 & 0.0441 & -0.0011 & -0.0087 \\ 
%   R & 0.3054 & 0.3560 & 0.2518 & 0.2438 \\ 
%   \hline
% \end{tabular}
% \caption{Difference in mean continuous ranked probability score averaged over all hold-out observations, and, in the case of reflectance (R), averaged over all wavelengths. The difference presented is independent minus joint. Positive values indicate that the joint model outperforms the independent model.}\label{tab:crps_comp}
% \end{table}
%\subsubsection{Model Adequacy}

\section{Results}\label{sec:results}

\subsection{Correlation Between Traits and Reflectance}\label{sec:cor_res}

\begin{figure}[h!]
    \centering
    \includegraphics[width = \textwidth]{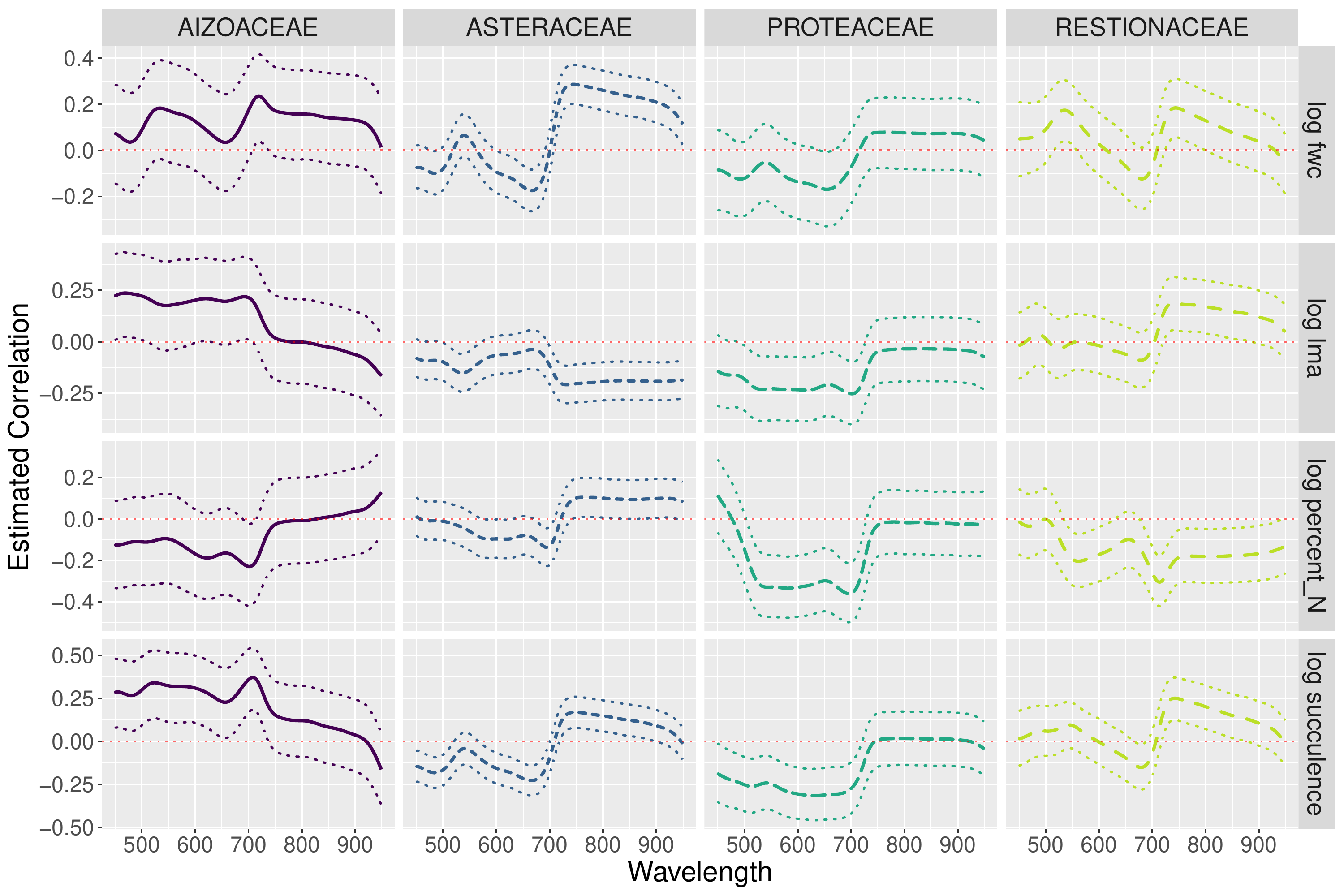}
    \caption{The estimated correlation between reflectance and all plant traits provided through $\bOmega$.}
    \label{fig:cor_trait_reflect}
\end{figure}

%In this subsection, we focus our discussion on the estimated correlations between the plant traits (log fresh water content, log leaf mass area, log percent Nitrogen, and log succulence) and the log reflectances. In Figure \ref{fig:cor_trait_reflect}, we plot the estimated correlations between log traits and reflectance for the Asteraceae, Proteaceae, and Restionaceae families. Specifically, we plot the posterior mean and 90\% credible intervals for between trait and reflectance correlation. 
We focus our discussion on the estimated correlations between the plant traits (log leaf water content, log leaf mass area, log percent Nitrogen, and log succulence) and the log reflectances. In Figure \ref{fig:cor_trait_reflect}, we plot the estimated correlations between log traits and reflectance for the Asteraceae, Proteaceae, and Restionaceae families. Specifically, we plot the posterior mean and 90\% credible intervals for between trait and reflectance correlation.

%\textbf{Henry, John - there is a lot of summary below -- is it useful? in terms of story, can you do anything with it?}
Given the environmental variables, we observe relatively weak relationships (typically between 0.2 to -0.2) between traits and reflectance. This finding is not entirely unexpected \citep{laughlin_intraspecific_2017, pau_poor_2022, wang_leaf_2022}, despite the amount of literature devoted to the strength of relationships and prediction potential between traits and reflectance within species and communities (see \cite{jacquemoud_extraction_2019} for a review). Fewer studies such as ours examine how these relationships vary across species within broader lineages such as families. The fact that we observe differing $\bT$ and $\bR$ relationships between families aligns with our expectation that the species within these families are comprised of various suites of leaf traits that in turn represent different adaptive strategies and ancestral constraints. In other words, spectra represent underlying biology, a point underscored in recent literature \citep{meireles_leaf_2020,cavender-bares_integrating_2022,kothari_plant_2022}. 

Further, we again highlight the fact that we are examining $\bT$ and $\bR$ relationships jointly given environment. This type of inquiry has often been done indirectly through controlled experimental manipulations \citep{thenot_photochemical_2002, inoue_relationship_2006, ripullone_effectiveness_2011, caturegli_effects_2020} or along environmental gradients \citep{coops_comparison_2002, asner_leaf_2009}. What our modelling shows is the explicit joint effects that environment has in terms of traits and reflectance. In cases of 0 or insignificant correlation we interpret that either the chosen environmental parameters have little effect on that lineage’s traits (which in turn affect reflectance) or that the species within the lineage differ in their responses to the same environmental parameter resulting in low signal, i.e., the ecological fallacy. In the case of significant correlations, we observe shifts in $\bT$ and $\bR$ relationships that suggest lineage-wide signals indicative of ecological and evolutionary processes such as adaptation or ancestral constraints.

%Overall, we observe relatively weak relationships between log reflectance and log fresh water content with one exception. For wavelengths greater than 700 nm (red and near infrared), we estimate that Asteraceae's reflectance is positively correlated with fresh water content; however, these correlations are small, estimated to be around 0.2.  On the other hand, Aizoaceae, Proteaceae, and Restionaceae show no or few wavelengths where the 90\% credible interval excludes 0. 

We observe relatively weak relationships between log reflectance and log leaf water content with one exception. For wavelengths greater than 700 nm (red and near-infrared [NIR]), we estimate that Asteraceae’s reflectance is positively correlated with leaf water content. The near infrared portion of the spectrum is well known in several instances to be a signal for various forms of water content in leaves \citep{pu_spectral_2003, rodriguez-perez_evaluation_2007, seelig_assessment_2008} and canopies \citep{penuelas_reflectance_1993, penuelas_estimation_1997}. On the other hand, Aizoaceae, Proteaceae, and Restionaceae show no or few wavelengths where the 90\% credible interval excludes 0. Note that this does not indicate that leaf water content and reflectance are unrelated in these lineages, but rather, after accounting for environment, these correlations are negligible. We suspect that the leaf water signal observed in the Asteraceae is attributable to the fact that the lineage has one of the broadest distributions extending from the arid Succulent Karoo to more mesic Fynbos.

%For log leaf mass area, some families exhibit positive correlations with reflectance, while others have negative correlations with reflectance. Both Asteraceae and Proteaceae show negative correlations between log LMA and log reflectance. Interestingly, Asteraceae shows negative correlations for wavelengths above 700 nm, while Proteaceae has negative correlations for wavelengths between 500 and 725 nm. Aizoaceae has positive estimated correlations between log LMA and reflectances for wavelengths less than 700 nm. On the whole, Restionaceae has very weak relationships between LMA and reflectance, but the 90\% credible intervals for wavelengths between 700 and 825 nm exclude 0. 

For log leaf mass per area, we generally expected to observe the strongest correlation within the near-infrared region (NIR), i.e., wavelengths greater than 700 nm \citep{asner_taxonomy_2011, jacquemoud_variation_2019, serbin_arctic_2019}.  Within the NIR, some families exhibit positive correlations with reflectance (Restionaceae), while others have negative correlations with reflectance (Asteraceae). Interestingly, Proteaceae has negative correlations for wavelengths between 500 and 725 nm for leaf mass per area. This could be the result of other co-correlated traits such as photosynthetic pigments that more often affect the visible region. There is some limited evidence within species for the negative correlation in the visible region for leaf mass per area \citep{ourcival_exploring_1999}.

%For Aizoaceae and Asteraceae, the relationship between log pN and reflectance appears to be weak for all wavelengths. Proteaceae shows negative correlations between log pN and reflectance for wavelengths between 500 and 725 nm. Although weak, the estimated relationship between pN and reflectance is negative for most wavelengths greater than 550 nm for Restionaceae.

Nitrogen within leaves is typically linked to the visible region of the spectrum (450-700 nm) given its strong links to the pigment chlorophyll, though there are nitrogen signals found at longer wavelengths \citep{jacquemoud_variations_2019}. For Aizoaceae and Asteraceae, the relationship between log pN and reflectance appears to be weak for all wavelengths, suggesting no lineage wide signals across the environmental range present in the study. Proteaceae shows negative correlations between log pN and reflectance for wavelengths between 500 and 725 nm. Although weak, the estimated relationship between pN and reflectance is negative for most wavelengths greater than 550 nm for Restionaceae.

%The relationship between log succulence and reflectance varies across families. Asteraceae has weak negative correlations between succulence and log reflectance for wavelengths less than 700 nm but weak positive correlations for wavelengths greater than 700 nm. Proteaceae shows negative correlations between succulence and log reflectance for wavelengths less than 725 nm, but the correlations are essentially 0 otherwise. As with LMA, succulence and reflectance have very weak relationships for Restionaceae. 

Leaf succulence is calculated by dividing the leaf area by leaf water content such that succulence represents the amount of water distributed throughout the leaf. Thus, we expected results like that for leaf water content. Overall, our results matched these expectations, but the results did have different significance compared to water content. Except for the Restionaceae, the other families had significant relationships within the visible range (wavelengths below 700 nm) which was surprising given the typically stronger signal of water within near infrared range. However, previous studies have found relationships between water content and the visible range attributed to the link between plant water status and photosynthetic machinery \citep{thenot_photochemical_2002, inoue_relationship_2006, ripullone_effectiveness_2011, hmimina_relationship_2014}. The Aizoaceae, a lineage dominated by succulent plants, now displayed significant, albeit weak, positive correlations for wavelengths below 700 nm. Asteraceae has weak negative correlations between succulence and log reflectance for wavelengths less than 700 nm but weak positive correlations for wavelengths greater than 700 nm. Proteaceae shows negative correlations between succulence and log reflectance for wavelengths less than 725 nm, but the correlations are essentially 0 otherwise. As with LMA, succulence and reflectance have very weak relationships for Restionaceae. 

%As discussed in Section \ref{sec:intro}, previous work has pointed out limited relationships between plant reflectances and plant traits. \textbf{Henry and John, can you add COMMENTS ABOUT THE SIGNIFICANCE OF THESE RELATIONSHIPS, even if weak.}

\subsection{Effects of the environmental and abundance predictors}\label{sec:res_env}

%In this subsection, we discuss the estimated effect of the environmental predictors and abundance on reflectance and traits. As discussed in Section \ref{sec:data}, because the scales of the covariates differ greatly, we have centered and scaled the covariates to aid in interpretation of the results. In Figure \ref{fig:coef_func}, we plot the four estimated coefficient functions and associated 90\% credible intervals for each family. In Figure \ref{fig:coef_T}, we plot the 90\% credible intervals for the regression coefficients for each trait. As we discuss, the effect of covariates varies greatly across families for both traits and reflectance. Although it is not always explicitly stated, all covariate effects should be interpreted as the estimated effect of a covariate holding all other covariates constant.  \textbf{Henry, John -- a lot of summaries, but not a lot of commentary here.  Hopefully, you can supplement!}

\begin{figure}[h!]
    \centering
    \includegraphics[width = \textwidth]{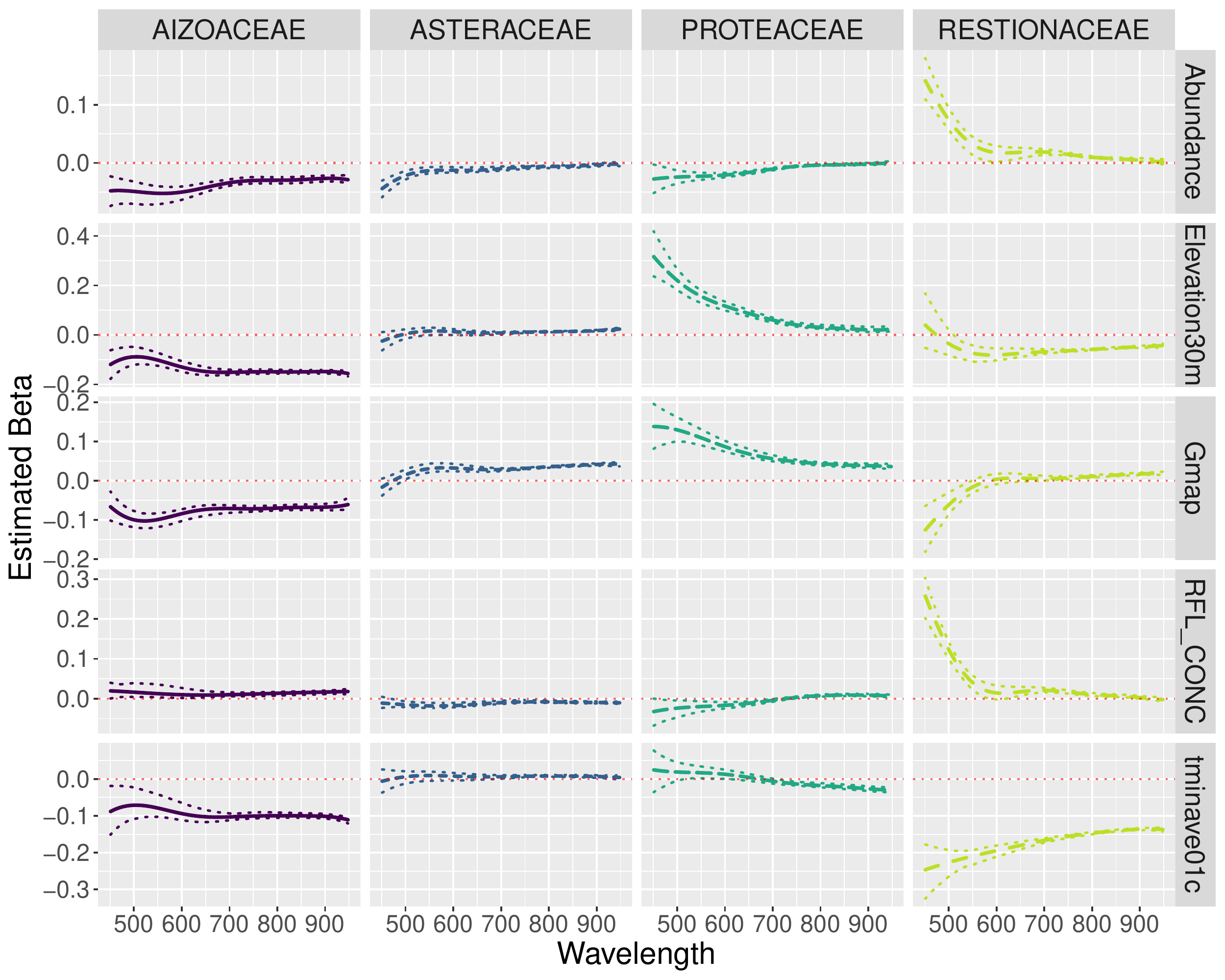}
    \caption{Estimated regression coefficient functions $\bbeta^{(R)}(w)$ the reflectance spectrum for all families and environmental predictors. }
    \label{fig:coef_func}
\end{figure}

\begin{figure}[h!]
    \centering
    \includegraphics[width = \textwidth]{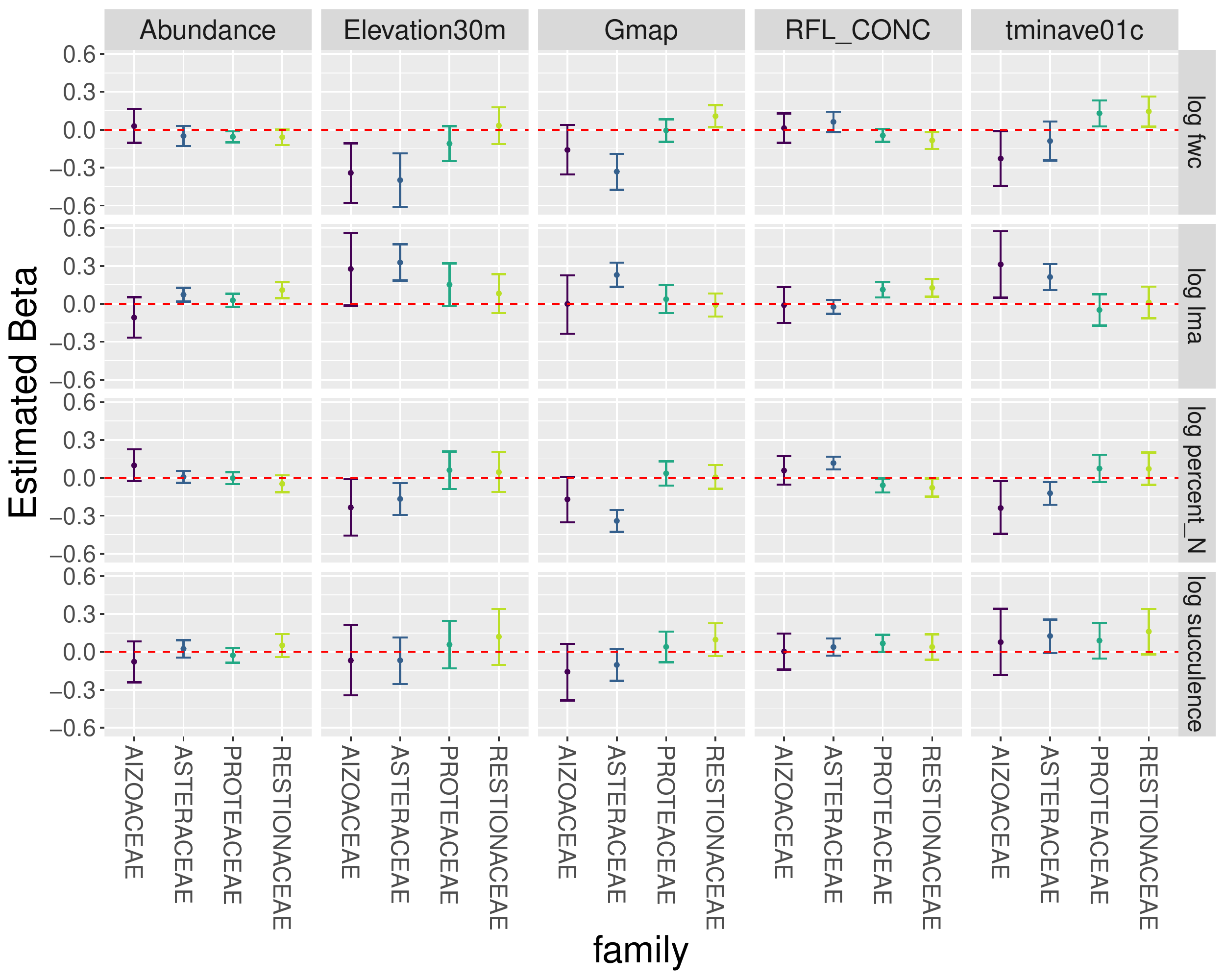}
    \caption{Estimated regression coefficients $\bbeta^{(T)}$ for all traits, families, and environmental predictors. }
    \label{fig:coef_T}
\end{figure}

We report the estimated effect of the environmental predictors and abundance on reflectance and traits. As discussed in Section \ref{sec:data}, we have centered and scaled the covariates to aid in interpretation of the results. In Figure \ref{fig:coef_func}, we plot the four estimated coefficient functions and associated $90\%$
credible intervals for each family. In Figure \ref{fig:coef_T}, we plot the $90\%$ credible intervals for the regression coefficients for each trait. Again, abundance is unlike the other environmental covariates in that it is not a direct driver of leaf traits and subsequent reflectances. 
%Rather we view this as an important variable to hold constant in comparison to other environmental co-variates so that we can downweigh rare species occurring at low abundances undue weight in the model. Further, abundance could be 
It is viewed as a proxy for unmeasured local environmental contributing to the “success” of species based on their biomass.

In terms of interpretation, we re-emphasize the fact that these models treat traits and reflectance jointly. As shown in Figure 3, the reflectance spectra within each lineage have clear correlations with environmental parameters. Our results in Figure \ref{fig:coef_func} are much weaker and less varied for all families, which was expected given the joint nature of the model where more variation is likely to be attributable to leaf traits \citep{laughlin_intraspecific_2017, pau_poor_2022, wang_leaf_2022}. We interpret Figure \ref{fig:coef_func} as capturing the signal of family-wide spectral responses to environment that are being driven by traits that are not currently measured, e.g., pigments, leaf surface features, or other measures of leaf anatomy. We would expect the trait and environment relationships in figure \ref{fig:coef_T} to roughly match the initial correlations in Figure \ref{fig:envr_EDA} given that it is traits that respond to environment and are subsequently manifested in the reflectance spectra. All covariate effects should be interpreted as the estimated effect of the particular environmental covariate holding all other environmental covariates constant.

\subsubsection{Aizoaceae}

%For Aizoaceae, reflectances have a negative relationship with abundance, elevation, annual precipitation and temperature for all wavelengths. On the other hand, Aizoaceae's reflectances have a weak positive relationship with rainfall concentration. Although the relationships are generally weak, it is the only family whose entire reflectance spectrum is related to all covariates. In terms of traits, elevation has a significant negative relationship with FWC and pN. Temperature has a positive relationship with LMA but negative relationships with FWC and pN. 

For Aizoaceae, reflectances have a negative relationship with abundance, elevation, annual precipitation and temperature for all wavelengths. In contrast, Aizoaceae’s reflectances have a very weak positive relationship with rainfall concentration. Although the relationships are generally weak, it is the only family whose entire reflectance spectrum is related to all covariates. We suspect that this could be a signal of overall leaf succulence given that precipitation and temperature have similar relationships and the expectation for leaves to be driven towards succulence as an adaptation to drier and hotter environments. Alternatively these family wide responses could be the result of other unmeasured traits such as those in leaf epidermal surfaces \citep{heim_effect_2015}.

 In terms of traits, elevation has a significant negative relationship with leaf water content (LWC) and percent nitrogen (pN). Temperature has a positive relationship with leaf mass per area (LMA) but negative relationships with LWC and pN. The LMA and temperature relationship roughly matches expectations found by the Leaf Economic Spectrum (LES), a study that examined trait and environment relationships at a global scale for a large sample of species \citep{wright_worldwide_2004}. The LES also finds that nitrogen and leaf mass per area tend to be negatively related, fitting in with our results as well. %We were surprised that leaf water decreased with temperature given that we would have expected a tendency towards greater water storage via leaf water storage in hotter environments. \textbf{Is this shooting ourselves in the foot?  Should we not mention it?}

\subsubsection{Asteraceae}

%For Asteraceae, the effects of all environmental covariates are very weak for the entire reflectance spectrum even though there are many significant estimated effects. Specifically, abundance and rainfall concentration are slightly negatively associated with log reflectance less than 700 nm, while annual precipitation has a weak positive relationship with reflectance for all wavelengths over 500 nm. 

%Fresh water content is negatively associated with elevation and annual precipitation, holding other covariates constant. On the other hand, leaf mass area is positively associated with elevation, precipitation, temperature, and abundance. Percent Nitrogen is negatively associated with elevation, annual precipitation, and January's minimum temperature, while it is positively associated with rainfall concentration.  Leaf succulence has a weak positive relationship with temperature, while all other covariates have 90\% credible intervals that include 0.

%\textbf{ADD COMMENTS}

For Asteraceae, the effects of all environmental covariates are very weak for the entire reflectance spectrum even though there are many significant estimated effects. Specifically, abundance and rainfall concentration are slightly negatively associated with log reflectance
less than 700 nm, while annual precipitation has a weak positive relationship with reflectance for all wavelengths over 500 nm. Out of the families chosen, the Asteraceae appear to have some of the weakest environmental signals for reflectance. This may be attributable to the fact that the Asteraceae is one of the most widely distributed groups across the Greater Cape Floristic Region, with a high diversity of growth forms, e.g., annuals, succulents and geophytes, tolerating a high number of environmental conditions \citep{manning_plants_2012, snijman_extra_2013}. This diversity would likely result in a high variation of reflectance signals that could weaken relationships.

LWC is negatively associated with elevation and annual precipitation, holding other covariates constant, while LMA is positively associated with elevation, precipitation, temperature, and abundance. The latter results have mixed correspondence to previous global analyses, with global LMA having negative to insignificant relationships with precipitation \citep{wright_worldwide_2004}.  Percent nitrogen is negatively associated with elevation, annual precipitation, and January’s minimum temperature, while it is positively associated with rainfall concentration. Leaf succulence has a weak positive relationship with temperature, while all other covariates have 90\% credible intervals that include 0. This partially matches the expectation that leaves would become succulent in more arid areas.
%though we have thought to also observe the same significance for annual precipitation. 

\subsubsection{Proteaceae}

%Of the four families, the log reflectance of Proteaceae has the strongest relationship with elevation and annual precipitation. For elevation, reflectance has a strong positive relationship for shorter wavelengths and a weak positive relationship for longer wavelengths. The estimated relationships between reflectance and annual precipitation is positive and significant for all wavelengths. On the other hand, reflectance has a weak negative relationship with temperature for wavelengths greater than 700 nm. Similarly, the relationship between abundance and reflectance is negative for wavelengths less than 750 nm. 

%In general, Proteaceae's traits have weak relationships with environmental covariates and abundance. However,  FWC reveals a slightly positive relationship with temperature and negative relationship with abundance while LMA has a slightly positive relationship with rainfall concentration.

Of the four families, the log reflectance of Proteaceae has the strongest relationship with elevation and annual precipitation. For elevation, reflectance has a strong positive relationship for shorter wavelengths and a weak positive relationship for longer wavelengths. As iterated previously, we interpret the environment and reflectance relationships in Figure \ref{fig:coef_func} as indicating trends in leaf traits that are unmeasured but varying across the environment. In the case of elevation and the visible region, this may be a trend in traits such as photosynthetic pigments which are strongly associated in the visible region of spectra \citep{jacquemoud_variation_2019}. The estimated relationships between reflectance and annual precipitation is positive and significant for all wavelengths, likely representing changes in traits co-correlated with the amount of water in leaves. Reflectance has a weak negative relationship with temperature for wavelengths greater than 700 nm. Similarly, the relationship between abundance and reflectance is negative for wavelengths less than 750 nm.

In general, Proteaceae’s traits have weak relationships with environmental covariates and abundance. However, LWC reveals a slightly positive relationship with temperature and negative relationship with abundance while LMA has a slightly positive relationship
with rainfall concentration. In a study of the genus \textit{Protea}, a prominent genus of the Proteaceae family within the Greater Cape Floristic Region, \cite{mitchell_functional_2015} found similar results for the LWC and temperature relationship (though theirs were non-significant) and LMA and rainfall seasonality (a related measure to rainfall concentration).
%\textbf{ADD COMMENTS}

\subsubsection{Restionaceae}

%Although Restionaceae showed a very weak correlations between traits and reflectance, reflectance is strongly connected with environmental covariates for shorter wavelengths. Abundance and rainfall concentration are positively related to reflectance at short wavelengths, while both annual precipitation and temperature are negatively related to reflectance. We estimate that increases in temperature are related to decreases in reflectance for all wavelengths. For wavelengths above 525 nm, we find a negative relationship between elevation and reflectance. 

%Turning to traits, Restionaceae's FWC is slightly positively related to temperature and annual precipitation but appears to be negatively related to abundance and rainfall concentration. Leaf mass area has slight positive relationships with rainfall concentration and abundance. For log pN and LS, all 90\% credible intervals include 0. 

%\textbf{ADD COMMENTS}

Although Restionaceae showed very weak correlations between traits and reflectance, reflectance is strongly connected with environmental covariates for shorter wavelengths, suggesting a shift in underlying traits associated with the visible region, e.g., photosynthetic pigments, along environmental gradients. Abundance and rainfall concentration are positively related to reflectance at short wavelengths, while both annual precipitation and temperature are negatively related to reflectance. We estimate that increases in temperature are related to decreases in reflectance for all wavelengths. For wavelengths above 525 nm, we find a negative relationship between elevation and reflectance.

Turning to traits, Restionaceae’s LWC is slightly positively related to temperature and annual precipitation but appears to be negatively related to abundance and rainfall concentration. Leaf mass area has slight positive relationships with rainfall concentration and
abundance. For log pN and LS, all 90\% credible intervals include 0. We suspect that the lack of clear trait and environmental trends could be a result of other known environmental drivers, e.g., fire and soil fertility, that drive differing adaptive strategies within the Restionaceae \citep{wuest_resprouter_2016}.

\section{Summary and Future Work}

%For investigation of leaf-scale trait levels and reflectances for plant families, we have presented modeling to enable assessment of the importance of environmental/habitat predictors in predicting traits and responses.  We have also taken up the useful objective of examining association between trait levels and reflectances across several families.  Lastly, we have shown that joint modeling of traits and reflectances provides better conditional predictive performance than modeling them independently.  We have illustrated this with data from four common families in the Greater Cape Floristic Region in South Africa.
For four plant families within the Greater Cape Floristic Region, we have presented modeling to enable assessment of the importance of environmental/habitat predictors in predicting traits and reflectance. This approach allows us to address the novel question of how trait and reflectance vary along environmental gradients. For remote sensing efforts aimed at regional and global extents, this question should be of immediate interest since it is the shifting nature of these relationships across different sets of plant functional types that reduces the generalizability of empirical models for trait prediction \citep{schimel_observing_2015, kothari_plant_2022, wang_leaf_2022}. Our current model presents an initial step in exploring an area that we feel has been under-utilized in ecology given a lack of available statistical tools. Lastly, we have shown that joint modeling of traits and reflectances provides better conditional predictive performance than modeling them independently.

In future work, our approaches could be adapted to include discrete or categorical traits, as in \cite{schliep2013multilevel} or \cite{clark2017generalized}. Extending the framework in \cite{white2021spatial} to spatially model the dependence between traits and reflectance would also be of interest, possibly including shape constraints \cite{white2021hierarchical}. In addition, with richer datasets, we could explore how reflectance/trait relationships vary along environmental gradients. Overall, the functional data approach in ecology is underutilized despite a plethora of ecological data that would be suitable for such analysis, e.g., spectral reflectance, organismal movement, and time series. We envision models incorporating joint responses of both scalar and functional data will be of high value to ecological problems beyond those in the present study.

%In future work, our approaches could be adapted to include discrete or categorical traits, as in \cite{schliep2013multilevel} or \cite{clark2017generalized}. In addition, extensions of the framework in \cite{white2021spatial} to spatially model the dependence between traits and reflectance may provide avenues for novel modeling and ecological inference.
%\textbf{SUGGESTIONS FOR FUTURE WORK?}
\backmatter

%\bmhead{Supplementary information}

\bmhead{Acknowledgments}

We thank Matthew Aiello-Lammens, Douglas Euston-Brown, Hayley
Kilroy Mollmann, Cory Merow, Jasper Slingsby, Helga van der Merwe, and Adam Wilson for their contributions in the data collection and curation. Special thanks to Cape Nature and the Northern Cape Department of Environment and Nature Conservation for permission for the collection of leaf spectra and traits. Data collection efforts were made possible by funding from National Science Foundation grant DEB-1046328 to J.A. Silander. Additional support was provided by NASA with a Future Investigators in NASA
Earth and Space Science and Technology (FINESST) grant award
(80NSSC20K1659) to H.A. Frye and J.A. Silander.

%\section*{Declarations}

% \appendix
\begin{appendices}

\section{Markov Chain Monte Carlo Details}\label{app:gibbs}
 
 In this section, we provide the full conditional distributions used for the Gibb's sampler. We use $\theta\vert \cdots$ to denote the full conditional distribution of the parameter $\theta$. For simplicity, we let $\bT$ be an $n \times s$ matrix of all observed traits and $\bR$ be an $n \times 500$ log reflectances. For traits and reflectances, respectively, we use $r^{(T)}_{-\theta}$ and $r^{(R)}_{-\theta}$ to be the residuals when excluding a the parameter $\theta$ ($\theta$ is used as a placeholder). For example, $r^{(R)}_{-\beta^{(R)}}$ is the residuals when removing the environmental regression from the model. In the case of wavelength-varying parameters, we use a similar notation to indicate the exclusion of the first term in a vector (e.g., $\theta_{-1}$). In addition, we let $D_{\sigma}$ be diagonal matrix with elements of $\sigma^2(w)$ and $D_{\beta^{(R)}}$ is a diagonal matrix with the prior variances given in \eqref{eq:priors}. In the case of updates for $\bU_j$, we refer to terms defined in Section \ref{sec:priors}.  The posterior conditional distributions are as follows:
 \begin{equation}
     \begin{aligned}
      \text{vec}(\bB^{(R)}) &\sim N(\Sigma_{\beta_R}\mu_{\beta_R},\Sigma_{\beta_R}) \\
 \text{vec}(\bB^{(T)}) &\sim N(\Sigma_{\beta_T}\mu_{\beta_T},\Sigma_{\beta_T}) \\
{\balpha^*}^{(R)} &\sim N(\Sigma_{\alpha_R}\mu_{\alpha_R},\Sigma_{\alpha_R}) \\
{\balpha}^{(T)}  &\sim N(\Sigma_{\alpha_T}\mu_{\alpha_T},\Sigma_{\alpha_T}) \\
     \end{aligned}
     \qquad
          \begin{aligned}
          \bU^{(R)}_j &\sim N(\Sigma_{U_j}\mu_{U_j},\Sigma_{U_j}) \\
          \sigma^{-2}_\alpha &\sim \text{Gamma}(a_{\alpha},b_{\alpha}) \\
\sigma^{-2}_{\beta_k} &\sim \text{Gamma}(a_{\beta_k},b_{\beta_k}) \\
\bOmega^{-1} &\sim \text{Wishart}\left(\nu_\Omega,S_\Omega \right)
     \end{aligned}
 \end{equation}
 
  \begin{equation}
     \begin{aligned}
     \mu_{\beta_R} &= (\bK_\beta' \otimes \bE') \text{vec}\left(r^{(R)}_{-\beta^{(R)}} D_{\sigma}^{-1} \right) \\
     \Sigma_{\beta_R} &= \left( \bK_\beta' D_{\sigma}^{-1}  \bK_\beta \otimes \bE' \bE + D_{\beta^{(R)}}^{-1} \right)^{-1}  \\
          \mu_{\beta_T} &= (\mathbb{I}_{s \times s} \otimes \bE') \text{vec}\left(r^{(T)}_{-\beta^{(T)}} {\Omega^{(T)}}^{-1} \right) \\
     \Sigma_{\beta_T} &= \left( {\Omega^{(T)}}^{-1} \otimes \bE' \bE+ D_{\beta^{(T)}}^{-1} \right)^{-1}  \\
          \mu_{\alpha_R} &= \bK_\alpha' \bone' r^{(R)}_{-\alpha^{(R)}} D_{\sigma}^{-1} \\
     \Sigma_{\alpha_R} &= \left( n \bK_\alpha' D_{\sigma}^{-1}  \bK_\alpha+ D_{\alpha^{(R)}}^{-1} \right)^{-1}  \\
               \mu_{\alpha_T} &= \bK_\alpha' \bone'  r^{(R)}_{-\alpha^{(T)}} {\Omega^{(T)}}^{-1} \\
     \Sigma_{\alpha_T} &= \left( n {\bOmega^{(T)}}^{-1} + D_{\alpha^{(T)}}^{-1} \right)^{-1}  \\
     \end{aligned}
     \qquad
          \begin{aligned}
          \mu_{U_j} &= \bK_U' D_{\sigma}^{-1} {r_j^{(R)}}_{-U_j} + \bSigma_{R_j \vert T_j}^{-1} \mu_{R_j\vert T_j} \\
          \Sigma_{U_j} &=\left( \bK_U'  D_{\sigma}^{-1} \bK_U + \bSigma_{R_j \vert T_j}^{-1} \right)^{-1} \\
     a_{\alpha} &= 1 + (N_\alpha -1) /2\\
    b_{\alpha} &= 1 + ({{\balpha^*_{-1}}^{(R)}}' {\balpha^*_{-1}}^{(R)}  ) /2\\ 
         a_{\beta_k} &= 1 + (N_\beta -1) /2\\
    b_{\beta_k} &= 1 + ({{\bB_{-1,k}}^{(R)}}' {{\bB_{-1,k}}^{(R)}} ) /2\\ 
    \nu_{\Omega} &= N_U + s + 1 + n \\
        S_{\Omega} &= \left( 10^{-3} \mathbb{I}  + \bU' \bU  \right)^{-1} \\
     \end{aligned}
 \end{equation}

\section{Model Comparison}\label{app:mod_comp}

In Tables \ref{tab:comp1} - \ref{tab:comp3}, we present the cross-validation results (in order) for Restionaceae, Proteaceae, and Aizoaceae. The joint model improves out-of-sample prediction performance for reflectances for all families, and this benefit is significant. On the other hand, the conditional out-of-sample prediction performance for traits depends on the family. For Asteraceae and Restionaceae, the families with the most data, prediction of plant traits benefits from joint modeling of traits and reflectance. For Proteaceae, plant trait predictions are slightly better under the independent model using the ES. However, we emphasize that the improved prediction is minimal for the independent model. For Aizoaceae, the family with the fewest data, prediction of plant traits suffers under the joint model. In summary, we find that reflectance predictions are uniformly and significantly better under the joint model for all plant families. Trait predictions are better under the joint model for two of the four families (in terms of ES) and only marginally worse for Proteaceae. We speculate that the benefit of the joint model appears when there is enough data to adequately estimate the relationship between traits and reflectance. Based on these findings, we use the joint model to present interpretation of the results.

\begin{table}[h!]
\centering
\caption{Model comparison between joint and independent models for Restionaceae.}\label{tab:comp1}
\begin{tabular}{|l|l|r|r|r|}
  \hline
Quantity & Model & MAE & RMSE & ES \\ 
  \hline  
log lma & $[T\vert E][R\vert E]$ & 0.354 & 0.431 & \multirow{4}{*}{0.585} \\ 
  log fwc & $[T\vert E][R\vert E]$ & 0.297 & 0.384 &  \\ 
  log succulence & $[T\vert E][R\vert E]$ & 0.443 & 0.580 &  \\ 
  log percent\_N & $[T\vert E][R\vert E]$ & 0.345 & 0.420 &  \\ 
    \hline  
  log Reflectance & $[T\vert E][R\vert E]$ & 0.587 & 1.079 & 17.925 \\ 
    \hline    \hline  
  log lma & $[T,R\vert E]$ & 0.294 & 0.424 &  \multirow{4}{*}{0.488} \\ 
  log fwc & $[T,R\vert E]$ & 0.239 & 0.315 &  \\ 
  log succulence & $[T,R\vert E]$ & 0.333 & 0.439 &  \\ 
  log percent\_N & $[T,R\vert E]$ & 0.306 & 0.532 &  \\ 
    \hline  
  log Reflectance & $[T,R\vert E]$ & 0.162 & 0.260 & 3.900 \\ 
   \hline
\end{tabular}

\end{table}

\begin{table}[h!]
\centering
\caption{Model comparison between joint and independent models for Proteaceae. }\label{tab:comp2}
\begin{tabular}{|l|l|r|r|r|}
  \hline
Quantity & Model & MAE & RMSE & ES \\ 
  \hline  
log lma & $[T\vert E][R\vert E]$ & 0.246 & 0.322 & \multirow{4}{*}{0.397} \\ 
  log fwc & $[T\vert E][R\vert E]$ & 0.186 & 0.274 &  \\ 
  log succulence & $[T\vert E][R\vert E]$ & 0.283 & 0.367 &  \\ 
  log percent\_N & $[T\vert E][R\vert E]$ & 0.217 & 0.287 & \\ 
  \hline
  log Reflectance & $[T\vert E][R\vert E]$ & 0.449 & 0.773 & 13.107 \\ 
      \hline \hline
  log lma & $[T,R\vert E]$ & 0.250 & 0.325 & \multirow{4}{*}{0.405} \\ 
  log fwc & $[T,R\vert E]$ & 0.215 & 0.303 &  \\ 
  log succulence & $[T,R\vert E]$ & 0.279 & 0.361 &  \\ 
  log percent\_N & $[T,R\vert E]$ & 0.238 & 0.303 &  \\ 
  \hline
  log Reflectance & $[T,R\vert E]$ & 0.139 & 0.224 & 3.432 \\ 
   \hline
\end{tabular}

\end{table}

\begin{table}[h!]
\centering
\caption{Model comparison between joint and independent models for Aizoaceae.}\label{tab:comp3}
\begin{tabular}{|l|l|r|r|r|}
  \hline
Quantity & Model & MAE & RMSE & ES \\ 
  \hline  
log lma & $[T\vert E][R\vert E]$ & 0.496 & 0.322 &\multirow{4}{*}{0.728 } \\ 
  log fwc & $[T\vert E][R\vert E]$ & 0.409 & 0.274 &  \\ 
  log succulence & $[T\vert E][R\vert E]$ & 0.490 & 0.367 &  \\ 
  log percent\_N & $[T\vert E][R\vert E]$ & 0.394 & 0.287 &  \\ 
    \hline
  log Reflectance & $[T\vert E][R\vert E]$ & 0.444 & 0.773 & 12.705 \\ 
    \hline \hline
  log lma & $[T,R\vert E]$ & 0.793 & 0.325 & \multirow{4}{*}{1.063}  \\ 
  log fwc & $[T,R\vert E]$ & 0.601 & 0.303 &  \\ 
  log succulence & $[T,R\vert E]$ & 0.445 & 0.361 &  \\ 
  log percent\_N & $[T,R\vert E]$ & 0.621 & 0.303 &  \\ 
    \hline
  log Reflectance & $[T,R\vert E]$ & 0.140 & 0.224 & 3.158 \\ 
   \hline
\end{tabular}

\end{table}

\end{appendices}

\bibliographystyle{apalike}
\bibliography{ref}

\end{document}